\documentclass[preprintnumbers,amsmath,amssymb,floatfix,prd,onecolumn,superscriptaddress,nofootinbib]{revtex4-1}
\usepackage{latexsym}
\usepackage{epsfig}
\usepackage{epstopdf}
\usepackage{amssymb}
\usepackage{mathtools}
\usepackage{amsmath,hyperref}
\usepackage{color}
\usepackage{graphicx}

\begin{document}
\title{Shadow and Deflection Angle of Rotating Black Holes in Perfect Fluid Dark Matter with a Cosmological Constant}
\author{Sumarna Haroon}\email{sumarna.haroon@sns.nust.edu.pk}\affiliation{Department of Mathematics, School of Natural
	Sciences (SNS), National University of Sciences and Technology
	(NUST), H-12, Islamabad, Pakistan}
\affiliation{Department of Physics and Astronomy, University of Waterloo, Waterloo, Ontario, Canada  N2L 3G1}
\author{Mubasher Jamil}
\email{mjamil@zjut.edu.cn}\affiliation{Department of Mathematics, School of Natural Sciences (SNS), National University of Sciences and Technology
(NUST), H-12, Islamabad, Pakistan}
\affiliation{Institute for Advanced Physics and Mathematics, Zhejiang University of Technology, Hangzhou, 310032, China}
\author{Kimet Jusufi} \email{kimet.jusufi@unite.edu.mk} \affiliation{Physics Department, State University of Tetovo, Ilinden Street nn, 1200,
Tetovo, Macedonia.}\affiliation{Institute of Physics, Faculty of Natural Sciences and Mathematics, Ss. Cyril
and Methodius University, Arhimedova 3, 1000 Skopje, Macedonia.}
\author{Kai Lin}\email{lk314159@hotmail.com}\affiliation{Institute of Geophysics and Geomatics, China University of Geosciences(Wuhan), Wuhan, Hubei 430074, China}
\affiliation{ Departamento de Ciencias e Ambientais, Escola de Engenharia de Lorena, Universidade de Sao Paulo, 12602-810, Lorena, SP, Brazil}
\author{Robert B. Mann} \email{rbmann@sciborg.uwaterloo.ca}\affiliation{Department of Physics and Astronomy, University of Waterloo, Waterloo, Ontario, Canada  N2L 3G1}

\begin{abstract}
 The presence of dark matter around a black hole remarkably affects its spacetime.
We consider the effects of dark matter on the shadow of a new   solution to the Einstein equations that describes a
 rotating black hole  in the background of perfect dark matter fluid (PFDM), along with its extension to nonzero
 cosmological constant $\Lambda$.   Working in  Boyer-Lindquist coordinates, we consider the effects of the PFDM parameter $\alpha$
 on the shadow cast by a black hole with respect to an observer at position $(r_o,\theta_o)$.
 By applying the Gauss-Bonnet theorem to the optical geometry
we find that notable distortions from a Kerr black hole can occur.  We describe their dependence on $\alpha$
and $\Lambda$.
\end{abstract}
\maketitle
\section{Introduction}

Astronomical predictions in recent years give us reason to expect that most  galaxies contain supermassive black holes at their  centre.  In particular, astrophysical observations strongly suggest the presence of a black hole, Sgr A*, at the centre of our galaxy. An interferometric instrument, {\it Gravity}, is actively working to obtain more precise observations of this supermassive black hole \cite{Gravity}. Simultaneously, a network of dishes all around the Earth has been developed using the VLBI technique to secure the shape and shadow of SgrA*. Known as the Event Horizon Telescope (EHT) \cite{EHT}, this project is  successfully collecting signals from radio sources, and it may be that we shall soon observe  the first silhouette of a super massive black hole. This data may also eventually provide us with a means for testing the general theory of relativity  in the strong-field regime.

It is therefore necessary to advance our theoretical research of black hole silhouettes (or shadows) to best evaluate the soon-expected observational data.
 Synge was the first to propose the apparent shape of a spherically symmetric black hole \cite{Synge}. After that Luminet \cite{Luminet} discussed the appearance of a Schwarzschild black hole surrounded by an accretion disk. The shadow of a Kerr black hole was first studied by Bardeen \cite{Bardeen}. Recent astrophysical advances have motivated many authors to invest in theoretical investigations of black hole shadows, including  Kerr-Newman black holes \cite{devaries}, naked singularities with deformation parameters \cite{hioki},   Kerr-Nut spacetimes \cite{abu} and more. The shadows of black holes in Chern-Simons modified gravity, Randall-Sundrum braneworlds, and
 Kaluza-Klein rotating black holes have been studied in \cite{Amarilla1,Amarilla2,Amarilla3}. Some authors have also tried to test theories of gravity by using the observations obtained from shadow of Sgr A* \cite{Bambi,Bro,B1,M}. A short review on shadows of a black hole is recently done in \cite{H1}.

%The idea is that we discuss the shadow of a Kerr anti-desitter/de Sitter black hole in the background of perfect fluid dark matter. This black hole metric is recently computed in \textbf{Xu, Z., Hou, X. and Wang, J., 2018. Kerr–anti-de Sitter/de Sitter black hole in perfect fluid dark matter background. Classical and Quantum Gravity, 35(11), p.115003.} In this black hole authors have computed Hamilton-Jacobi equation but that is when cosmological constant is zero, but we are taking a general case.
%In this paper we study the effect of perfect fluid dark matter parameter $\alpha$ on the shadow of a rotating black hole. To do so the manuscript is section wise partitioned as: Section (I) is a short review of the Kerr AdS/dS black hole in presence of perfect fluid dark matter.

The method used for computing shadows of (rotating) black holes  is more or less the same in all cases. An observer is placed at a very large distance (effectively infinity) away from the black hole, and it is from the viewpoint of this observer that the shadow is determined; typically celestial coordinates are introduced.  More generally, the
 bending of light in a given spacetime background is a result of the spacetime curvature due to the presence of a massive body, and the deflection for a given impact parameter is obtained by solving the geodesic equations. Another geometric method for computing the deflection of light  \cite{Gibbons} involves integration over a domain outside the light ray using the Gauss-Bonnet theorem. This method really shows the  global aspect of the lensing effect in terms of the topology of the spacetime. Subsequently this method was applied to study lensing in different black hole/wormhole geometries \cite{j1,j2,j3,j4,j5,j6,j7,j8,gab1,gab2}.

For asymptotically flat black holes these methods are  fine, but  in the presence of a cosmological constant there is an additional subtlety in that the position of the observer needs to be fixed.  While the effect of the cosmological constant on the deflection of light has been investigated by several authors, unfortunately there seems to be no general agreement on the final results \cite{cs1,cs2,cs3,cs4,cs5,cs6,cs7,cs8}.  The main difficulty relies on the fact that the main assumption  according to which the source and observer are located at infinity is no longer valid in the case of  non-asymptotically flat spacetimes.

The  Standard model of cosmology suggests that our universe is compiled of $27\%$ dark matter and 68\% dark energy, while the rest is baryonic matter. Though dark matter has not been directly detected,   observational evidence for its existence can be found in abundance. Examples include galactic rotation curves \cite{Rubin}, the dynamics of galaxy clusters \cite{Zwicky}, and the measurements of cosmic microwave background anisotropies obtained through PLANCK \cite{Planck} .

It is therefore natural to ask how black hole solutions might depend on perfect fluid dark matter.  Recently a generalization of the Kerr-(A)dS solution in the presence of dark matter (PFDM) was obtained \cite{Perfect}.  This solution had a number of interesting features.  The size of its ergosphere decreased with increasing $|\alpha|$, where $\alpha$ parameterizes the strength of the dark matter contribution to the metric.  Null circular stable orbits were shown to exist, and
the dependence of the rotational velocity on $\alpha$ was determined.  However no observational consequences of this solution were considered.

  Motivated by the above we investigate here the deflection of light and the shadow of
the rotating PDFM black hole \cite{Perfect}. We note that the shadow of black hole in the presence of quintessence \cite{Pratap,Abu} and in a dark matter halo \cite{Zhou} has been previously considered.
We intend here to use  techniques recently employed  \cite{Grenz} for computing the shadow of a rotating black hole with cosmological constant.    We begin by  fixing the location of the observer in Boyer-Lindquist coordinates $(r_0,\theta_0)$, the respective radial and polar angular coordinates of the observer. Instead of considering photon rays coming from the past,  we follow them from the location $(r_0,\theta_0)$ to the past. The behaviour of such light-like geodesics can be characterized into two categories: those that venture so close  to the outer horizon $r=r_+$ of the black hole  that they are absorbed by it due to the gravitational pull, and those that ultimately escape to
their original source in the past.   Thus a boundary is defined, between these two categories of light-like geodesics, which encloses a dark region called the \emph{shadow}.

 Despite the continued debate over the effects of the cosmological constant,  we shall follow  recent work by Ishihara et al. \cite{ish1,ish2,ish3,ish4} that  takes into consideration finite-distance corrections in two particular spacetimes:  Schwarzschild-de Sitter spacetime and an exact solution in Weyl conformal gravity \cite{ish1}. To address issues with non-asymptotically flat spacetimes we shall consider the effects of the PFDM parameter $\alpha$ and  cosmological constant on the deflection angle assuming finite distance corrections.

\section{Black Holes in Perfect Fluid Dark Matter Background}

 Amongst the many dark matter models that have been suggested is   the perfect fluid dark matter model, which was initially proposed by Kiselev \cite{K1}, and entailed construction of a new class of spherically symmetric black hole  metrics  in the presence of PFDM {\cite{Li}}.  In the spherically symmetric case this class of black holes was distinguished by
a new term in the metric function that grows logarithmically with distance from the black hole.   The logarithmic dependence was introduced by Kiselev \cite{K1} to account for the asymptotic behaviour of the
quintessential matter at large distances, i.e. in the halo dominated region, in order to explain the asymptotic rotation curves for the dark matter (see also \cite{Li}).
Only recently has this class been generalized to include rotation \cite{Perfect}, providing a PFDM version of the Kerr-(A)dS solution.
The metric is given by
\begin{eqnarray}
ds^2&=&-\frac{\Delta_r}{\Xi^2\Sigma}\left(dt-a\sin^2\theta{d\phi}\right)^2+\frac{\Delta_\theta\sin^2\theta}{\Xi^2\Sigma}\left(adt-(r^2+a^2)d\phi\right)^2\\\label{metric}
&+&\frac{\Sigma}{\Delta_r}dr^2+\frac{\Sigma}{\Delta_\theta}d\theta^2,
\end{eqnarray}
where
\begin{eqnarray}\nonumber
\Delta_r&=&r^2-2Mr+a^2-\frac{\Lambda}{3}r^2\left(r^2+a^2\right)+\alpha r \log\frac{r}{|\alpha|},\\\label{values}
\Delta_\theta&=&1+\frac{\Lambda}{3}a^2\cos^2\theta, \qquad\text{and}\qquad{\Xi=1+\frac{\Lambda}{3}a^2}, \qquad{ \Sigma=  {r^2 + a^2\cos^2\theta}},
\end{eqnarray}
with the mass parameter of the black hole being $M$. The parameter indicating the presence of perfect fluid dark matter is $\alpha$. This solution reduces to a rotating black hole in a PFDM background when $\Lambda=0$, and to the Kerr-(A)dS solution for $\alpha=0$.   The PDFM stress-energy tensor in the standard orthogonal basis of the Kerr-(A)dS metric can be written in diagonal form $[\rho,p_r,p_\theta,p_\phi]$, where
\begin{eqnarray}
\rho = -p_r = \frac{\alpha r}{8\pi \Sigma^2}, \qquad  p_\theta = p_\phi = \frac{\alpha r}{8\pi \Sigma^2}\left(r - \frac{\Sigma}{2r}\right).
\end {eqnarray}

For $\Lambda\neq 0$, the solution can either be a Kerr-Anti-de Sitter ($\Lambda < 0$) or Kerr-de Sitter ($\Lambda > 0$) metric. The horizons of the black hole are the solutions of $\Delta_r=0$ i.e.
\begin{eqnarray}\label{horizon}
\frac{\Lambda}{3}r^4+\left(\frac{\Lambda}{3}a^2-1\right)r^2+2Mr-a^2+\alpha r\log\left(\frac{r}{\mid\alpha\mid}\right)=0.
\end{eqnarray}
In general there are inner and outer horizons for Kerr and Kerr anti-de Sitter black holes,
with an additional cosmological horizon for Kerr-de Sitter black holes. Imposing the requirement that PFDM does not change the number of horizons
as compared to its Kerr counterpart, the parameter $\alpha$ is constrained such that \cite{Perfect}
\begin{eqnarray}
\alpha\in
 \begin{cases}
  (-7.18M,0)\cup(0,2M)\quad\text{if}\quad \Lambda =0, \\
   (\alpha_{min},0)\cup (0,\alpha_{max})\quad\text{if}\quad \Lambda\neq{0}
  \end{cases}
\end{eqnarray}
where $\alpha_{max}$ and $\alpha_{min}$ respectively satisfy
\begin{eqnarray}
\alpha_{min}&+&\alpha_{min}\log(\frac{2M}{-\alpha_{min}})=2M+H(\Lambda),\\  \nonumber
\alpha_{max}&+&\alpha_{max}\log(\frac{2M}{\alpha_{max}})=2M+H(\Lambda),
\end{eqnarray}
and
\begin{eqnarray}
H(\Lambda)= - \textrm{sgn}(\Lambda) \left(\frac{32}{3\Lambda}M^3+\frac{2}{3}\Lambda a^2\right),
\end{eqnarray}
and we see if $a=0$ that $H > 0$ for   $\Lambda <0$ and  $H < 0$ for   $\Lambda > 0$.

\section{Photon Region}

For the spacetime (\ref{metric}),  geodesic motion is governed by the Hamilton Jacobi equation \cite{Chandra}:
\begin{eqnarray}\label{hj}
-\frac{\partial{S}}{\partial{\tau}}=\frac{1}{2}g^{\mu\nu}\frac{\partial{S}}{\partial{x}^\mu}\frac{\partial{S}}{\partial{x^\nu}},
\end{eqnarray}
where $\tau$ is an affine parameter, $x^\mu$ represents the four-vector $(t,r,\theta,\phi)$ and $S$ is  Hamilton's principal function, which can be made separable by introducing an ansatz such that
\begin{eqnarray*}
S=\frac{1}{2}\delta\tau-Et+L\phi+S_r(r)+S_\theta(\theta),
\end{eqnarray*}
where energy $E$ and angular momentum $L$ are constants of motion related to the associated Killing vectors
$\partial/\partial t$ and $\partial/\partial \phi$.
 For timelike geodesics $\delta=1$ and for null geodesics $\delta=0$.
%\begin{eqnarray}\label{hj1}
%-\delta\Sigma=-\frac{\Xi^2}{\Delta_r}\left((r^2+a^2)E-aL\right)^2+\frac{l^2}{\Delta_\theta\sin^2 \theta}\left(a\sin^2\theta{E}-L\right)^2+\Delta_r\left(\frac{\partial S_r}{\partial r}\right)^2+\Delta_\theta\left(\frac{\partial S_\theta}{\partial\theta}\right)^2
%\end{eqnarray}
%Separating the above equation in radial and angular parts
%\begin{eqnarray}\nonumber
%\Delta_r^2\left(\frac{dS_r}{dr}\right)^2&=&\Xi^2\left(\left(r^2+a^2\right)E-aL\right)^2-\Delta_rr^2\delta-\mathcal{C}\Delta_r=R(r),\\\label{separate}
%\Delta_\theta^2\left(\frac{dS_\theta}{dr}\right)^2&=&-\frac{\Xi^2}{\sin^2\theta}\left(aE\sin^2\theta-L\right)^2-a^2\delta\cos^2\theta+\mathcal{C}\Delta_\theta=\Theta(\theta).
%\end{eqnarray}
%The solution for $S$ is now
%\begin{eqnarray}
%S=\frac{1}{2}\delta\tau-Et+L\phi+\int^{r}\frac{\sqrt{R(r)}}{\Delta_r}dr+\int^{\theta}\frac{\sqrt{\Theta(\theta)}}{\Delta_\theta}d\theta
%\end{eqnarray}
%where $S_r$ and $S_\theta$ are given by
%\begin{eqnarray}
%S_r=\int\frac{\sqrt{R}}{\Delta_r}dr\quad~\text{and}\quad~S_\theta=\int\frac{\sqrt{\Theta}}{\Delta_\theta}
%\end{eqnarray}
Thus by solving Eq. (\ref{hj}) the resulting equations describing the propagation of a particle are
\begin{eqnarray}\label{tdot}
\Sigma\dot{t}&=&\Xi^2\frac{\left((r^2+a^2)E-aL\right)(r^2+a^2)}{\Delta_r}-\frac{a\Xi^2\left(aE\sin^2\theta-L\right)}{\Delta_\theta},\\\label{rdot}
\Sigma^2\dot{r}^2&=& \Xi^2\left(\left(r^2+a^2\right)E-aL\right)^2-\Delta_rr^2\delta-\mathcal{C}\Delta_r=R(r),\\\label{thdot}
\Sigma^2\dot{\theta}^2&=&-\frac{\Xi^2}{\sin^2\theta}\left(aE\sin^2\theta-L\right)^2-a^2\delta\cos^2\theta+\mathcal{C}\Delta_\theta=\Theta(\theta),\\\label{phidot}
\Sigma\dot{\phi}&=&\frac{a\Xi^2\left((r^2+a^2)E-aL\right)}{\Delta_r}-\frac{\Xi^2\left(aE\sin^2\theta-L\right)}{\Delta_\theta\sin^2\theta},
\end{eqnarray}
  for both null and time-like geodesics. In the above equations, besides the two constants of motion $E$ and $L$, we also have the Carter constant $\mathcal{C}$ \cite{Carter}.
    As we are interested in black hole shadows, henceforth  we consider only  null geodesics, for which $\delta=0$. To reduce the number of parameters we write $\xi=L/E$ and $\eta=\mathcal{C}/E^2$, and  rescale $R/E^2 \to R$ and $\Theta/E^2 \to \Theta$. Then Eq. (\ref{rdot}) and (\ref{thdot}) respectively yield
\begin{eqnarray}\label{R}
R&=&\Xi^2\left((r^2+a^2)-a\xi\right)^2-\Delta_r\eta,
\end{eqnarray}
and
\begin{eqnarray}\label{Theta}
\Theta &=&\eta\Delta_\theta-\frac{\Xi^2}{\sin^2\theta}\left(a\sin^2\theta-\xi\right)^2.
\end{eqnarray}

The photon region is defined as the region of space where gravity is strong enough that photons are forced to travel in orbits.
Circular photon orbits  only exist in the equatorial plane for rotating Kerr black holes, and there are two such types,  retrograde and prograde. To this end, we note that there are other solutions such as the rotating dyonic black holes in Kaluza-Klein and Einstein-Maxwell-dilaton theory, for which circular photon orbits do not exist on the equatorial plane \cite{pedro}. Note that Schwarzschild is another counter-example, albeit static, that contains non-equatorial circular photon orbits due to spherical symmetry. To determine the photon region  we  require $r=r_{s}$ such that
\begin{eqnarray}\label{cond1}
R(r_s)=0\quad~ \text{and}\quad~ \left.\frac{dR(r)}{dr}\right|_{r=r_s}  =0,
\end{eqnarray}
along with the condition that
\begin{eqnarray}\label{cond2}
\Theta(\theta)\geq 0\quad~\text{for}\quad~\theta\in{[0,\pi]}
\end{eqnarray}
By  solving   (\ref{cond1}) we obtain the value of $\xi$ and $\eta$ at $r=r_s$ to be
\begin{eqnarray}\label{xi}
a\xi(r_s)&=&r_s^2+a^2-4r_s\frac{\Delta_r(r_s)}{\Delta_r^\prime(r_s)},\\
\label{eta}
\eta(r_s)&=&\frac{16r_s^2\Xi^2\Delta_r(r_s)}{\left(\Delta_r^\prime(r_s)\right)^2},
\end{eqnarray}
and by inserting Eqs. (\ref{xi}) and (\ref{eta}) in condition (\ref{cond2}) we find the condition
\begin{eqnarray}\label{phregion}
\left(4r\Delta_{r_s}-\Sigma\Delta_{r_s}^\prime\right)^2\leq16a^2r^2\Xi^2\Delta_{r_s}\Delta_\theta\sin^2\theta,
\end{eqnarray}
that describes the photon region. For $\Lambda=\alpha=0$, Eq. \eqref{phregion} yields in the equatorial plane
the Kerr result $r = 2m\left(1+\cos\left(\frac{2}{3}\cos^{-1}\left(\pm \frac{|a|}{m}\right)\right)\right)$.
Photon orbits can be stable or unstable. The unstable photon orbit at $r=r_s$ exists when $\frac{d^2R(r_s)}{dr^2}>0$, which also defines the boundary of the black hole shadow. Thus the positive solution of
\begin{eqnarray}\label{stable}
\frac{R^{\prime\prime}(r_s)}{8E^2\Xi^2} = r_s^2+2r_s\Delta_{r_s}{\Delta_{r_s}}^\prime-2r_s^2\frac{\Delta_{r_s}\Delta_{r_s}^{\prime\prime}}{(\Delta_{r_s}^\prime)^2},
\end{eqnarray}
determines the contour of the shadow. Here we have restored the factor of $E$ and $\prime$ denotes the derivative with respect to $r$.

\section{Shadows of the Kerr PFDM Black Hole}

As noted above, in the presence of a cosmological constant the position of the observer needs to be fixed, employing the technique recently introduced in \cite{Grenz}. So we fix the observer in Boyer-Lindquist coordinates $(r_0,\theta_0)$, where $r_0$ is the radial coordinate and $\theta_0$ is angular coordinate of observer. We also assume that the observer is in domain of outer communication i.e. $\Delta_r>0$ and we consider the trajectories of light rays sent from position $(r_0,\theta_0)$ to the past.

We now define orthonormal tetrads $(e_0,e_1,e_2,e_3)$ at the observer's position $(r_0,\theta_0)$ such that
\begin{eqnarray}\label{e1}
e_0&=&\frac{\Xi^2}{\sqrt{\Delta_r\Sigma}}\left(\left(r^2+a^2\right)\partial_t+a\partial_\phi\right)\biggr\rvert_{(r_0,\theta_0)},\\\label{e2}
e_1&=&\sqrt{\frac{\Delta_\theta}{\Sigma}}\partial_\theta\biggr\rvert_{(r_0,\theta_0)},\\\label{e3}
e_2&=&-\frac{\Xi^2}{\sqrt{\Delta_\theta\Sigma }\sin\theta}\left(\partial_\phi+a\sin^2\theta\partial_t\right)\biggr\rvert_{(r_0,\theta_0)},\\\label{e4}
e_3&=&-\sqrt{\frac{\Delta_r}{\Sigma}}\partial_r\biggr\rvert_{(r_0,\theta_0)},
\end{eqnarray}
where $e_0$ is observer's four velocity, $e_0\pm e_3$ are tangent to the direction of principal null congruences and $e_3$ is along the spatial direction pointing towards the centre of the black hole. Let the coordinates of the light ray are described as $\lambda(s)=(r(s),\theta(s),\phi(s),t(s))$, then a vector tangent to $\lambda(s)$ is given by
\begin{eqnarray}\label{vector1}
\dot{\lambda}=\dot{r}\partial_r+\dot{\theta}\partial_\theta+\dot{\phi}\partial_\phi+\dot{t}\partial_t.
\end{eqnarray}
This tangent vector can also be described in terms of orthonormal tetrads and celestial coordinates $\rho$ and $\sigma$ as
\begin{eqnarray}\label{vector2}
\dot{\lambda}=\beta\left( -e_0+\sin\rho\cos\sigma e_1+\sin\rho\sin\sigma e_2+\cos\rho e_3\right),
\end{eqnarray}
where the scalar factor $\beta$ is obtained from Eq. (\ref{vector1}) and (\ref{vector2}) such that
\begin{eqnarray}
\beta=g(\dot{\lambda},e_0)=\left.\Xi^2\frac{aL-E(r^2+a^2)}{\sqrt{\Delta_r\Sigma}}\right\vert_{(r_0,\theta_0)}.
\end{eqnarray}
Our next aim is to define the celestial coordinates, $\rho$ and $\sigma$ in terms of parameters $\xi$ and $\eta$. To do so we compare the coefficients of $\partial_r$ and $\partial_\phi$ in Eq. (\ref{vector1}) and (\ref{vector2})
and thus we obtain
\begin{figure}[h!]
\includegraphics[width=0.4\textwidth]{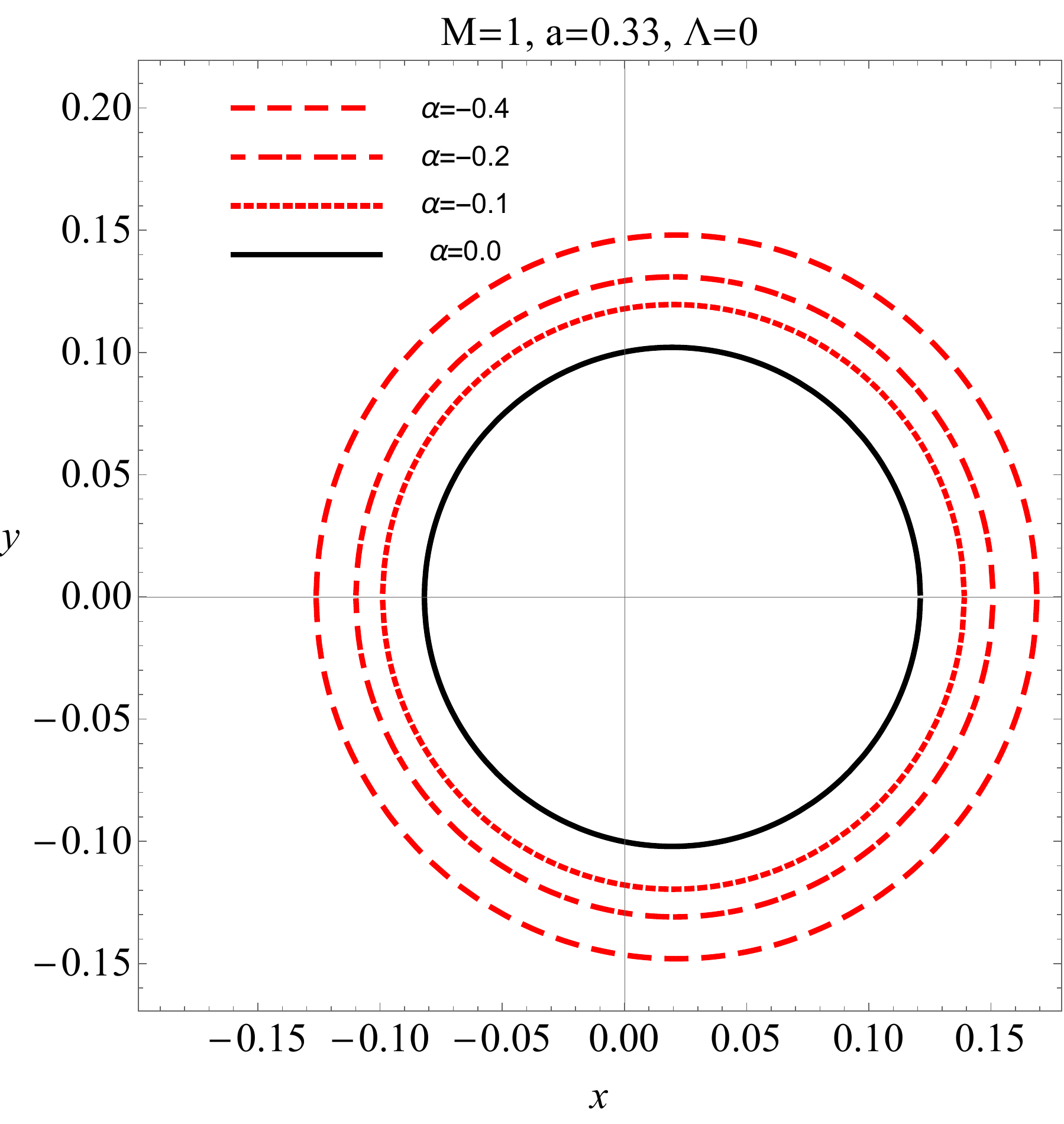}
\includegraphics[width=0.44\textwidth]{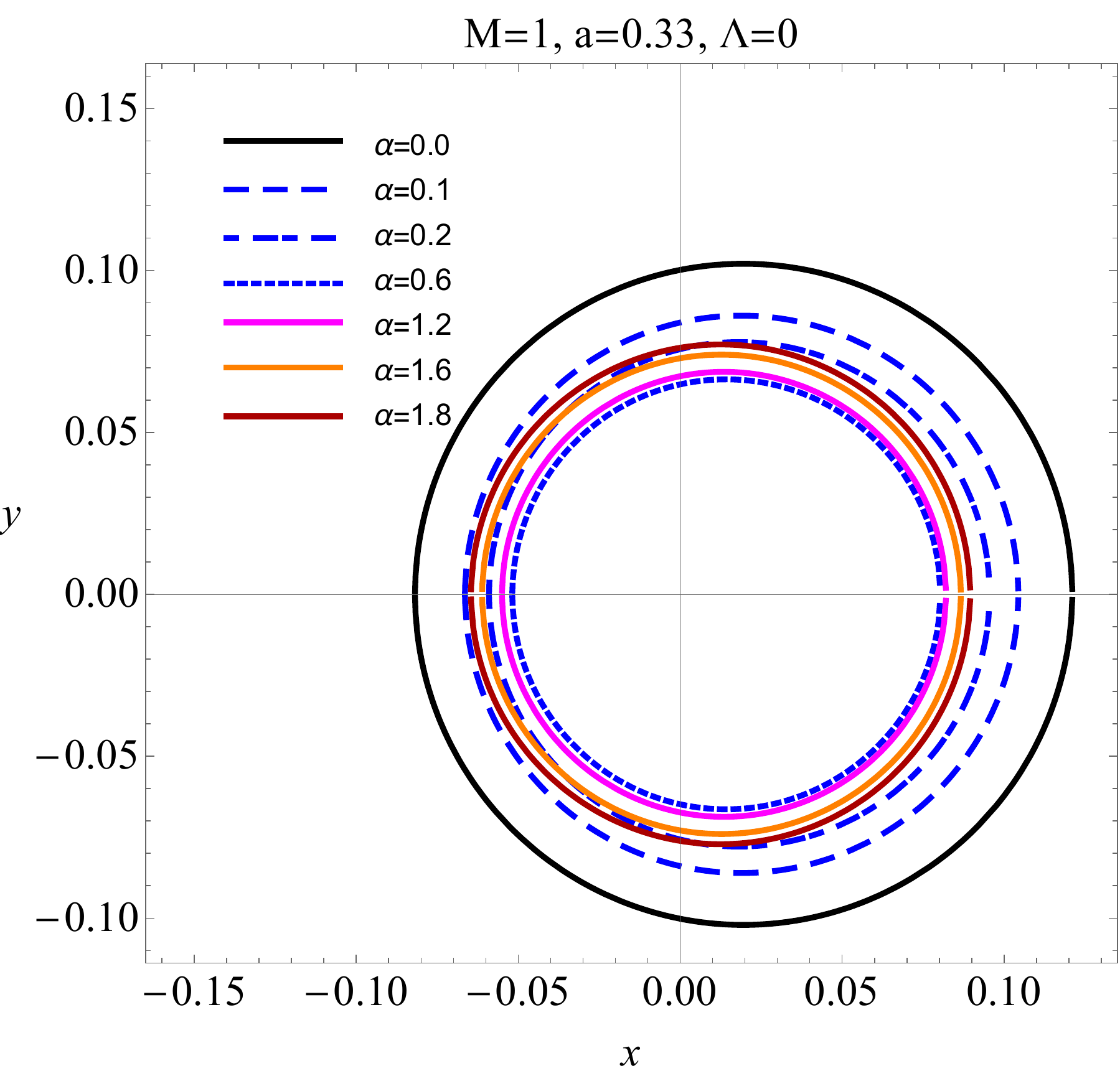}\\
\includegraphics[width=0.4\textwidth]{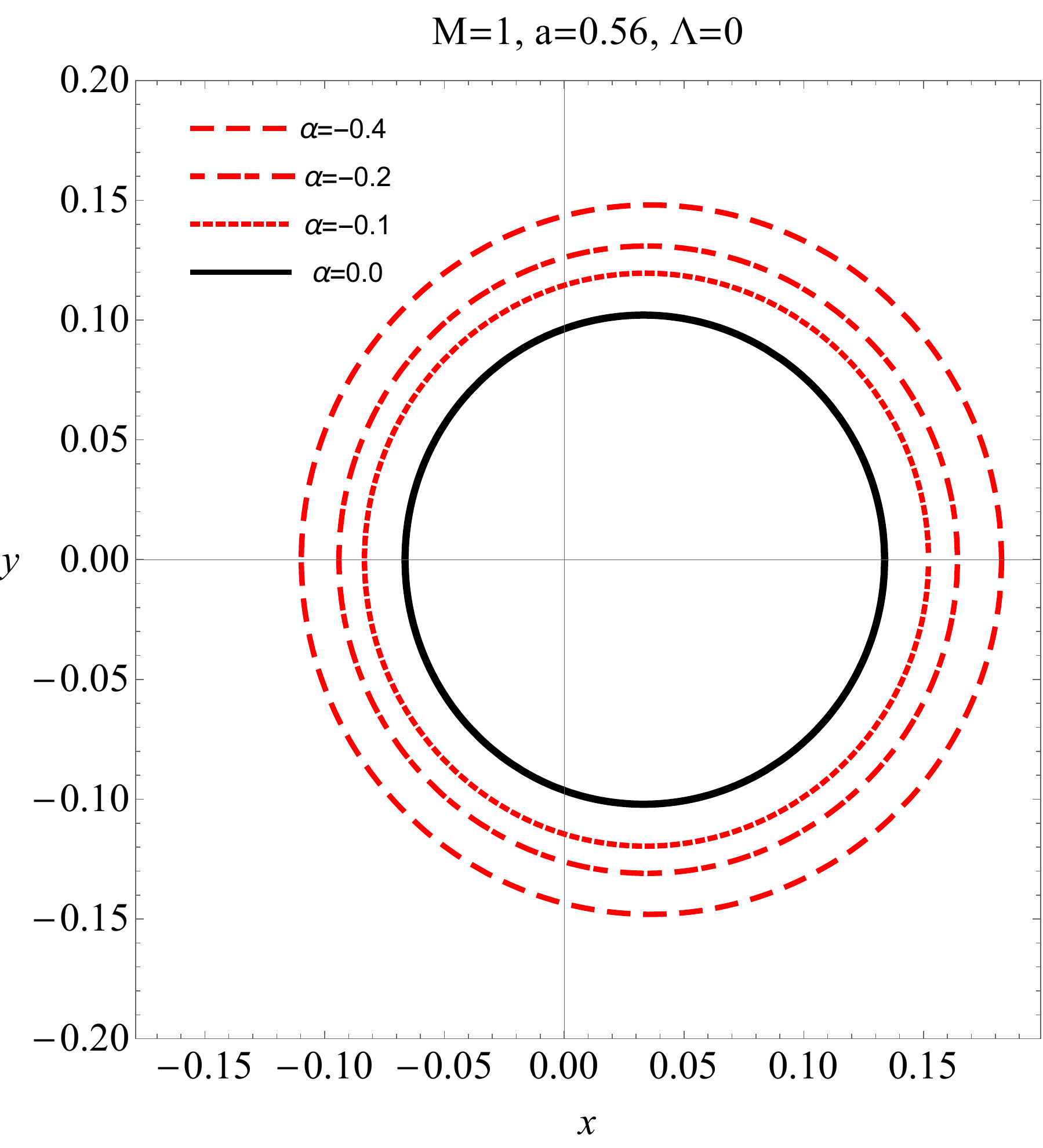}
\includegraphics[width=0.42\textwidth]{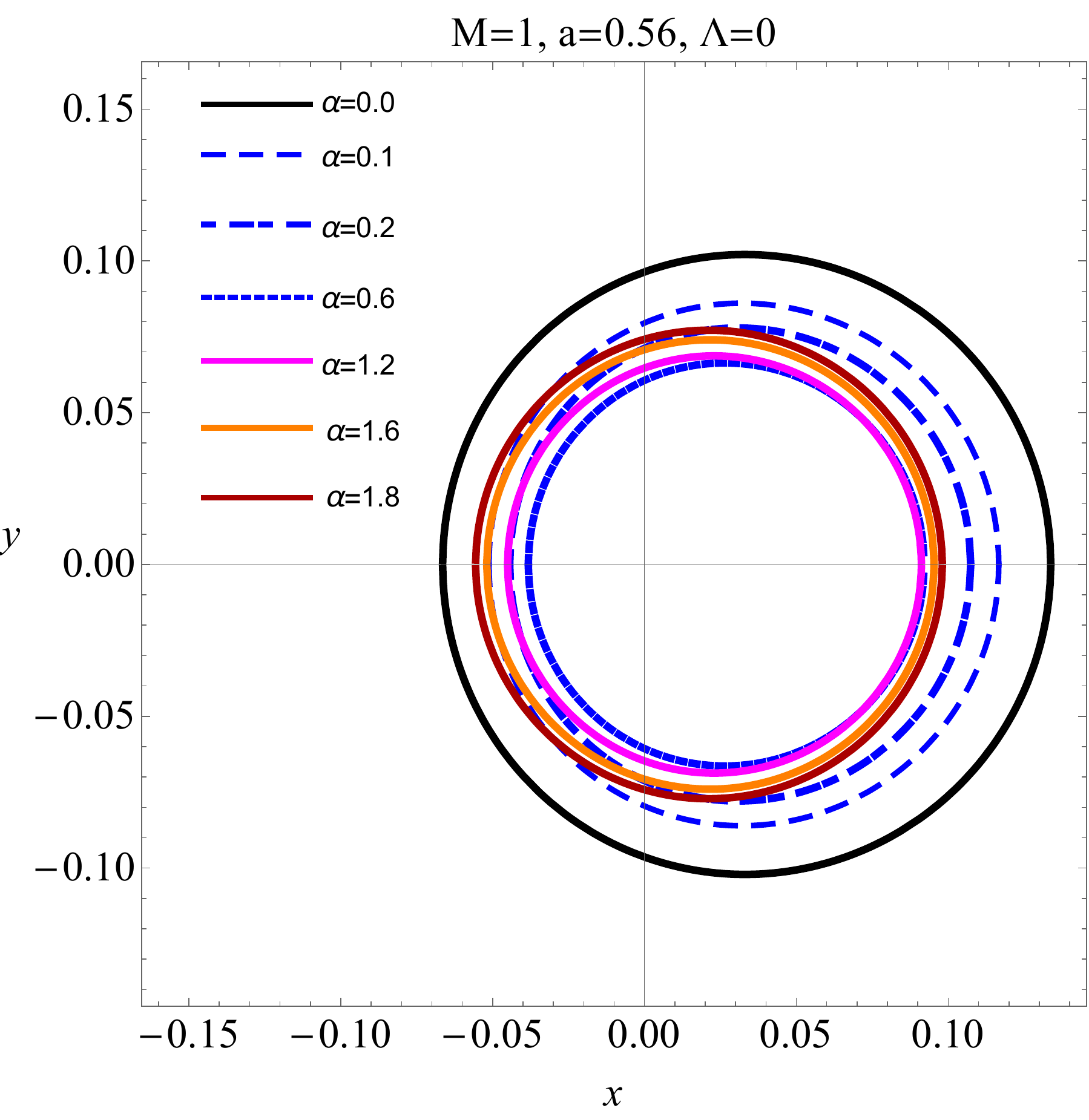}\\
\includegraphics[width=0.42\textwidth]{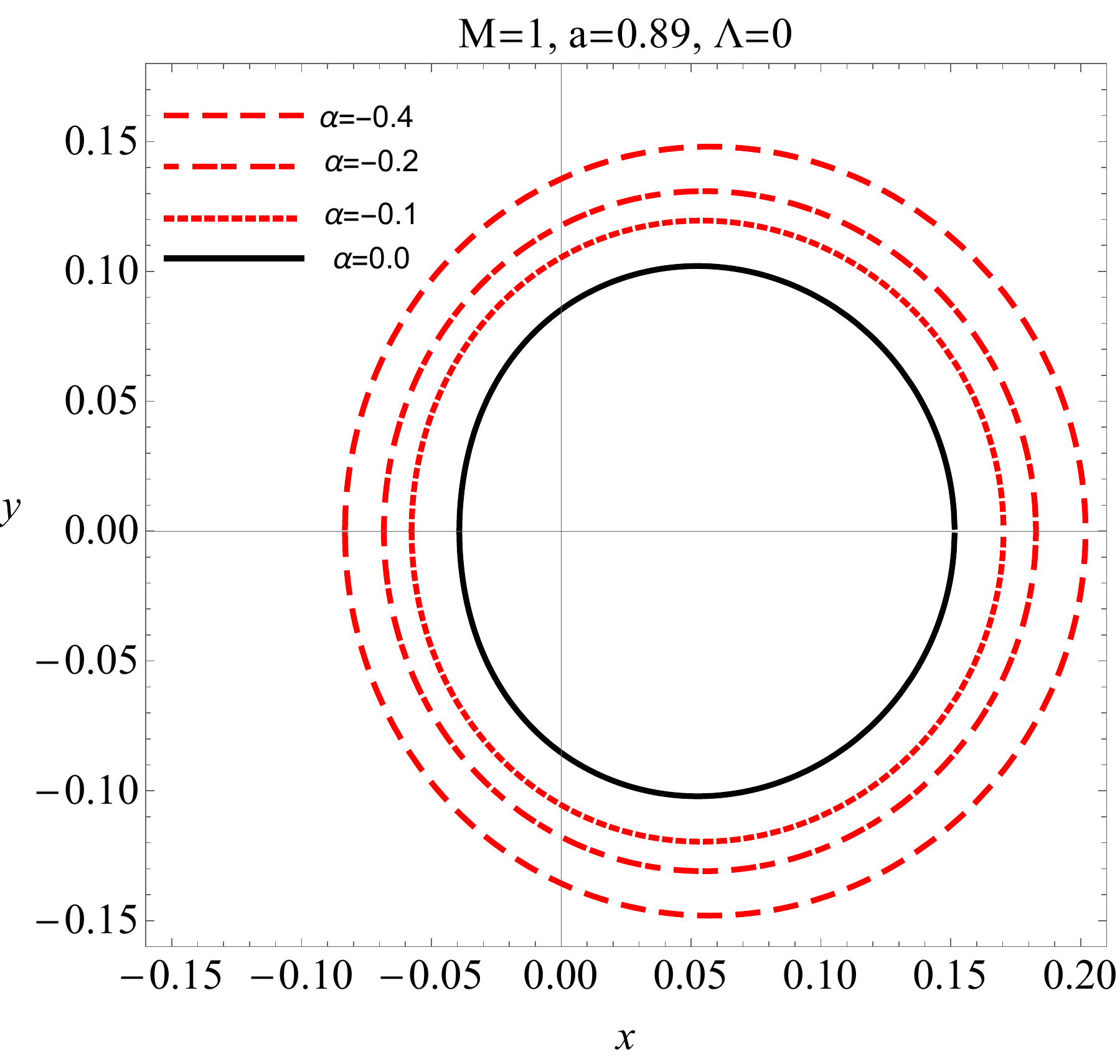}
\includegraphics[width=0.41\textwidth]{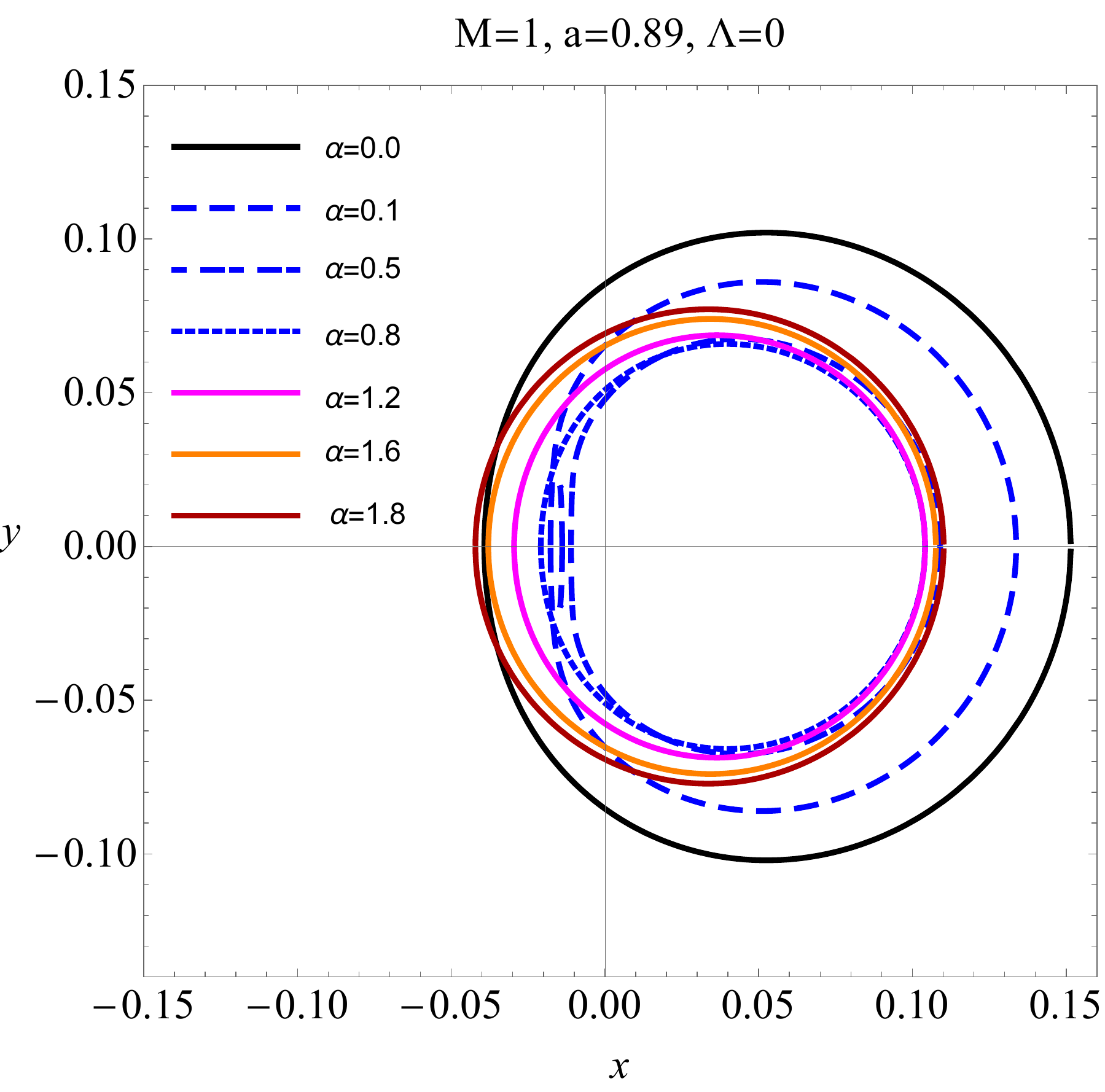}
 \caption{Shadows cast by a rotating black hole in PFDM background for different values of $\alpha$; all quantities
 are in units of $M$. The observer is positioned at $r_0=50$ and $\theta_0=\pi/2$.
 %$\alpha=-0.8$ (Dashed),$-1.0$ (DotDashed), $ -1.2$(Dotted), $1.0$ (\textcolor{red}{Dashed}), $2.8$ (\textcolor{red}{DotDashed}) and $4.0$ (\textcolor{red}{Dotted}). The value of cosmological constant is $\Lambda=-10^{-4}$ (upper panel), $\Lambda=0$ (middle panel) and $\Lambda=10^{-7}$ (lower panel).
     }\label{Shadow1}
     \end{figure}

\begin{eqnarray}\label{rho}
\sin\rho=\left.\sqrt{1-\frac{\dot{r}^2\Sigma^2}{\Xi^4\left(\left(r^2+a^2\right)E-aL\right)^2}}\right\vert_{(r_0,\theta_0)}.
\end{eqnarray}
and
\begin{eqnarray}\label{sigma}
\sin\sigma=\left.{\frac{\sqrt{\Delta_\theta}\sin\theta}{\sqrt{\Delta_r}\sin\rho}}\left(\frac{\Sigma\Delta_r}{\Xi^2\left(\left(r^2+a^2\right)E-aL\right)}\dot{\phi}-a\right)\right\vert_{(r_0,\theta_0)}.
\end{eqnarray}
Using Eqs. (\ref{rdot}) and ({\ref{phidot}}), we can present the above two equations in terms of parameter $\xi$ and $\eta$ as
\begin{eqnarray}
\sin\rho=\left.\frac{\pm\sqrt{\Xi^2\left(\Xi^2-1\right)\left(\left(r^2+a^2\right)-a\xi\right)^2+\Delta_r\eta}}{\Xi^2\left(r^2+a^2-a\xi\right)}\right\vert_{(r_0,\theta_0)},
\end{eqnarray}
and
\begin{eqnarray}
\sin\sigma=\left.\frac{\sqrt{\Delta_r}\sin\theta}{\sqrt{\Delta_\theta}\sin\rho}\left[\frac{a-\xi\csc^2\theta}{a\xi-(r^2+a^2)}\right]\right\vert_{(r_0,\theta_0)}.
\end{eqnarray}
The boundary of shadow of the black hole can be presented graphically by projecting a stereographic projection from the celestial sphere onto to a plane with the Cartesian coordinates
\begin{eqnarray}
x=-2 \tan\left(\frac{\rho}{2}\right)\sin(\sigma), \\
y=-2 \tan\left(\frac{\rho}{2}\right)\cos(\sigma).
\end{eqnarray}

 Figure \ref{Shadow1} allows us to distinguish the silhouette cast by a rotating black hole in presence of perfect fluid dark matter $(\alpha\neq 0)$ from that of Kerr black hole $(\alpha=0)$.   For $\alpha < 0$ we find that
 the shadow of the black hole gets larger and more circular as $\alpha$ becomes increasingly negative.  However
 for $\alpha > 0$ the effect on the shadow is no longer monotonic. For small $\alpha > 0$ the shadow shrinks whilst maintaining its asymmetric shape.  However once $\alpha \gtrsim 0.8$, the shadow begins to grow, becoming increasingly circular and shifting leftward relative to its $\alpha=0$ Kerr counterpart.

 Our study thus indicates that presence of perfect fluid dark matter  can have considerable effects on a black hole silhouette. The rotational distortion of a Kerr black hole is diminished for sufficiently large $|\alpha|$, even for
large spin $(a=0.84)$.  The next effect is that the PFDM   `cancels out' the rotational distortion of the shadow.

\begin{figure}[h!]
   \includegraphics[width=8cm]{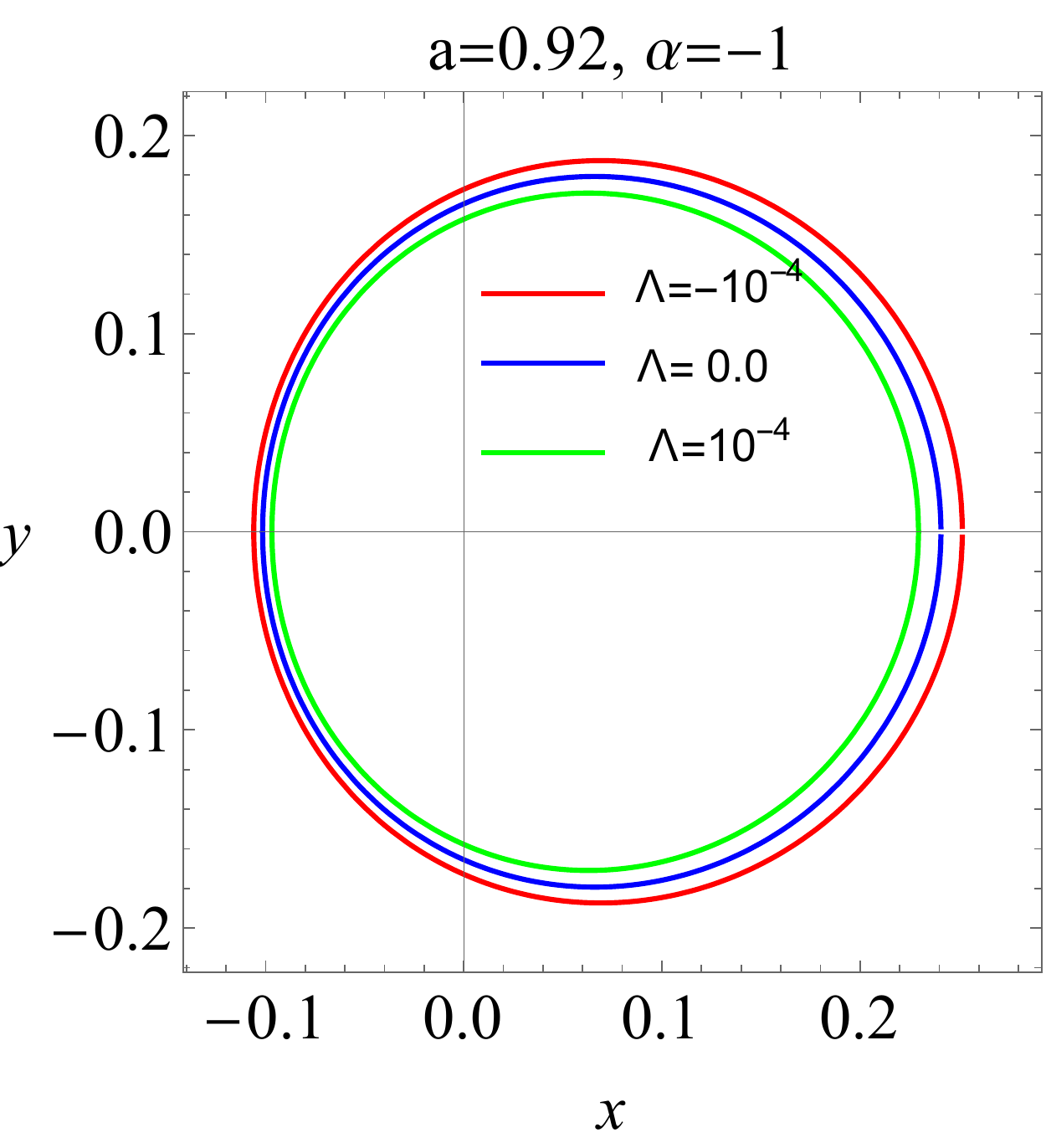}
  \includegraphics[width=8cm]{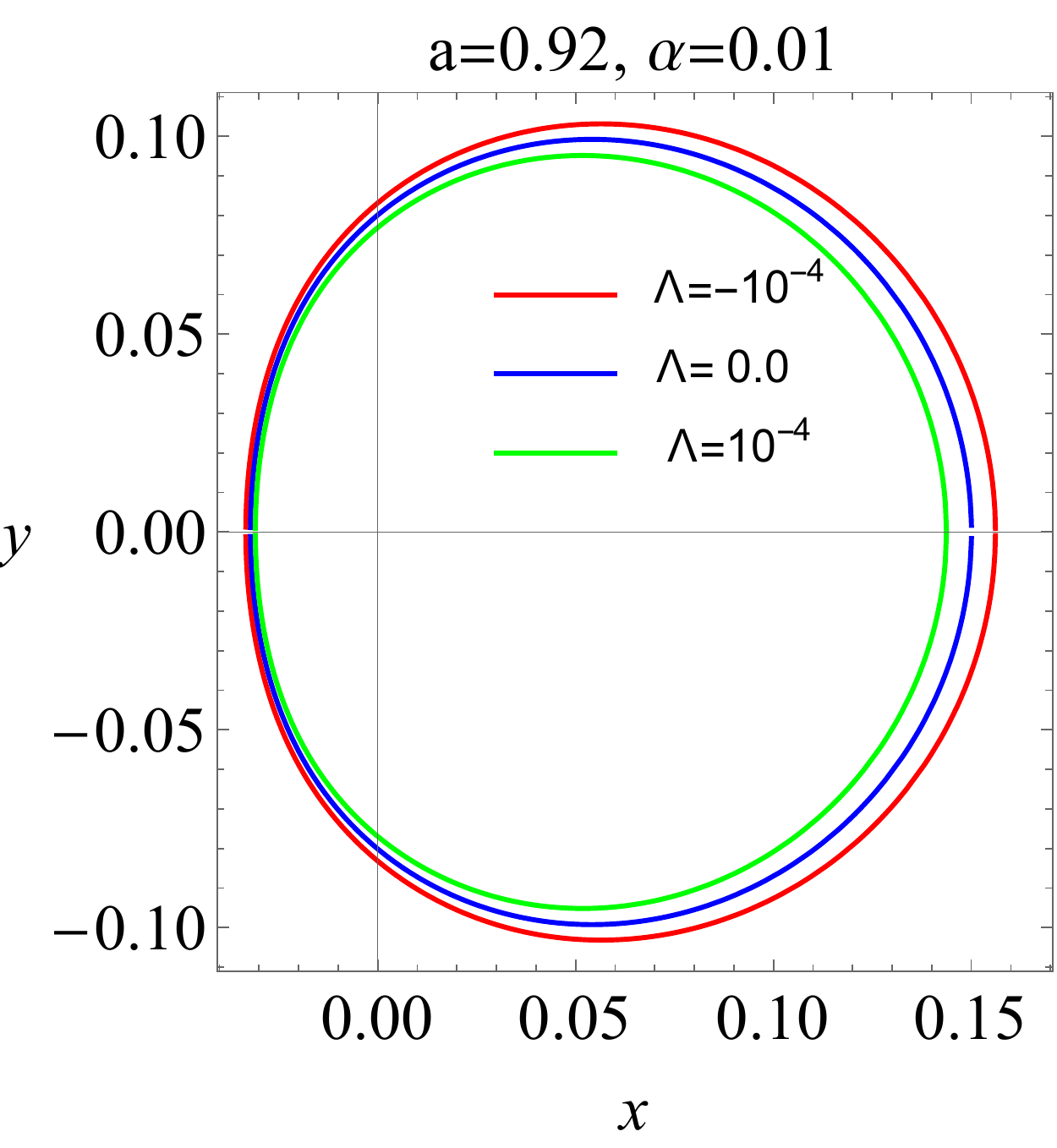}\\
  \includegraphics[width=8cm]{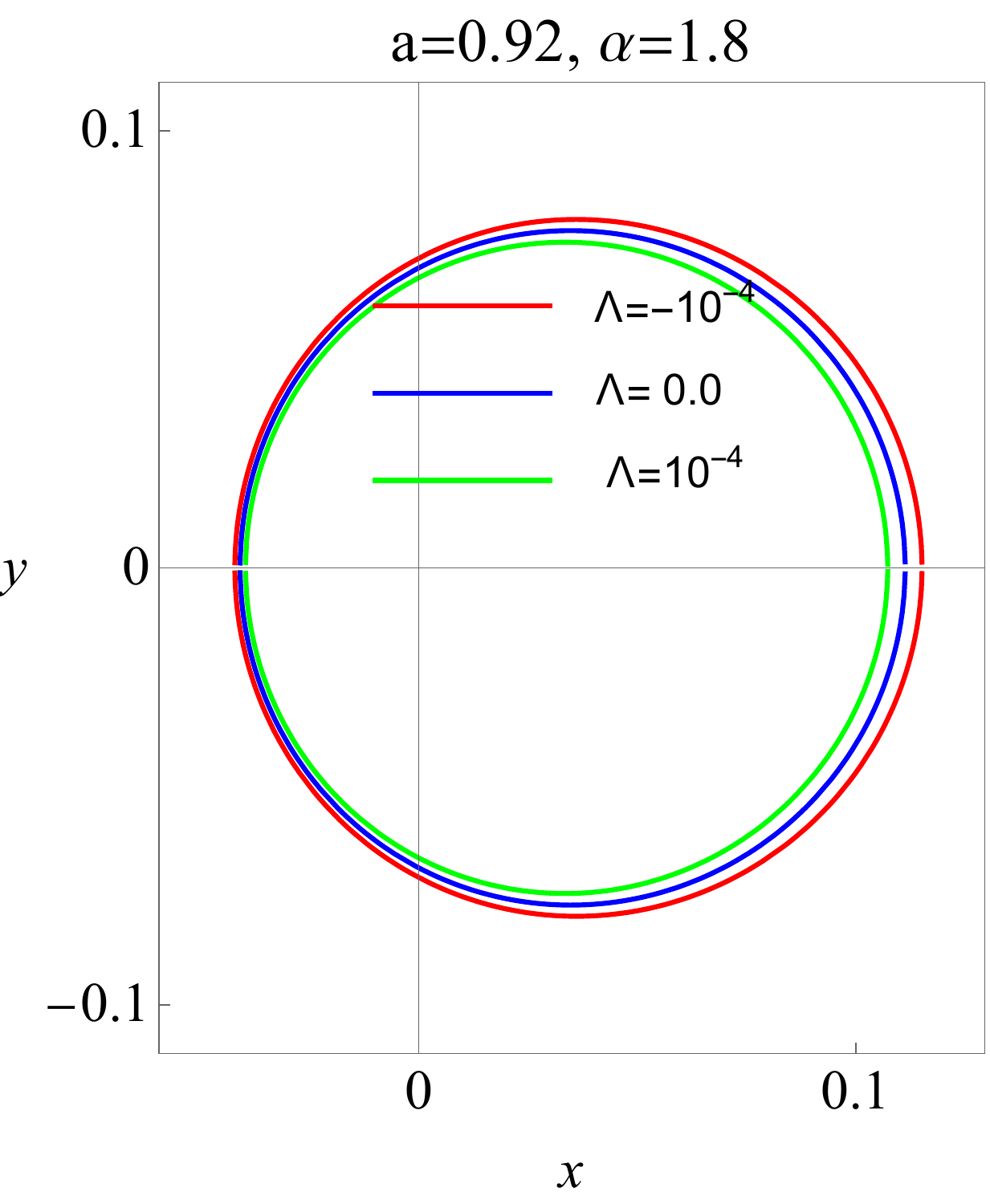}
\caption{Variation in shadow of a rotating black hole in PFDM background w.r.t cosmological constant, when the observer is at position $r_0=50$ and $\theta_0=\pi/2$.
All quantities are in units of $M$.
}\label{Shadow2}
 \end{figure}

 Figure \ref{Shadow2} shows the effects of cosmological constant on the shadow for different values of parameter $\alpha$.  We see that for small $|\Lambda|$ the shadow maintains its shape for a given $\alpha$, increasing for
 the AdS case $\Lambda < 0$ and decreasing for the dS case $\Lambda > 0$.

\section{Deflection of light}

\subsection{Deflection angle without a cosmological constant}

In this section we proceed to study the deflection angle of light applying  the Gauss Bonnet Theorem (GBT)  over the optical geometry under the  assumption that the distance from the  source $(S)$ to the  receiver $(R)$ is finite.  In order to see more clearly the effect of the PFDM parameter and the cosmological constant on the deflection of light first consider  $\Lambda=0$.

Let $T$ be a two-dimensional orientable surface with boundaries $\partial T_a (a = 1, 2, \ldots, N)$, and let the jump angles between
the curves be $\theta_a (a = 1, 2, \ldots, N).$ In terms of this construction the GBT can be stated as follows \cite{ish1,ish2}
\begin{equation}
\iint_T \mathcal{K} dS+\sum_{a=1}^{N}\int_{\partial T_a} \kappa_g dl+\sum_{a=1}^{N}\theta_a=2 \pi ,
\end{equation}
in which $\mathcal{K}$ is the Gaussian curvature of the surface $T$, $dS$ gives the surface area element, $\kappa_g$ is known as the geodesic curvature of $\partial T_a$, and finally $l$ is the line element along the boundary.   It is convenient to find first the black hole optical metric by imposing the null condition $ds^2=0$, and then by solving the spacetime  metric for $dt$, yielding the generic form
\begin{eqnarray}
dt= {\pm}\sqrt{\gamma_{ij}dx^i dx^j}+\beta_i dx^i,
\end{eqnarray}
where $i,j$ run from $1$ to $3$. Furthermore $\gamma_{ij}$ and $\beta_i$ in our case are found as follows
\begin{eqnarray}\notag
\gamma_{ij}dx^i dx^j&=& \left(\frac{r^4}{\Delta_r (\Delta_r-a^2 \sin^2\theta)}   \right)dr^2+ \left(\frac{r^4}{\Delta_r-a^2 \sin^2\theta}   \right)d\theta^2 \\
&+&  \left(\frac{\sin^2\theta  \Delta_r (r^2+\cos^2 \theta a^2)^2}{(\Delta_r-a^2 \sin^2\theta)^2}   \right)d\phi^2,\\
\beta_i dx^i &=& -\frac{ \sin^2 \theta \,(a^2+r^2-\Delta_r)a}{\Delta_r-a^2 \sin^2 \theta} d\phi.
\end{eqnarray}

\begin{figure}[h!]
 \includegraphics[width=0.60\textwidth]{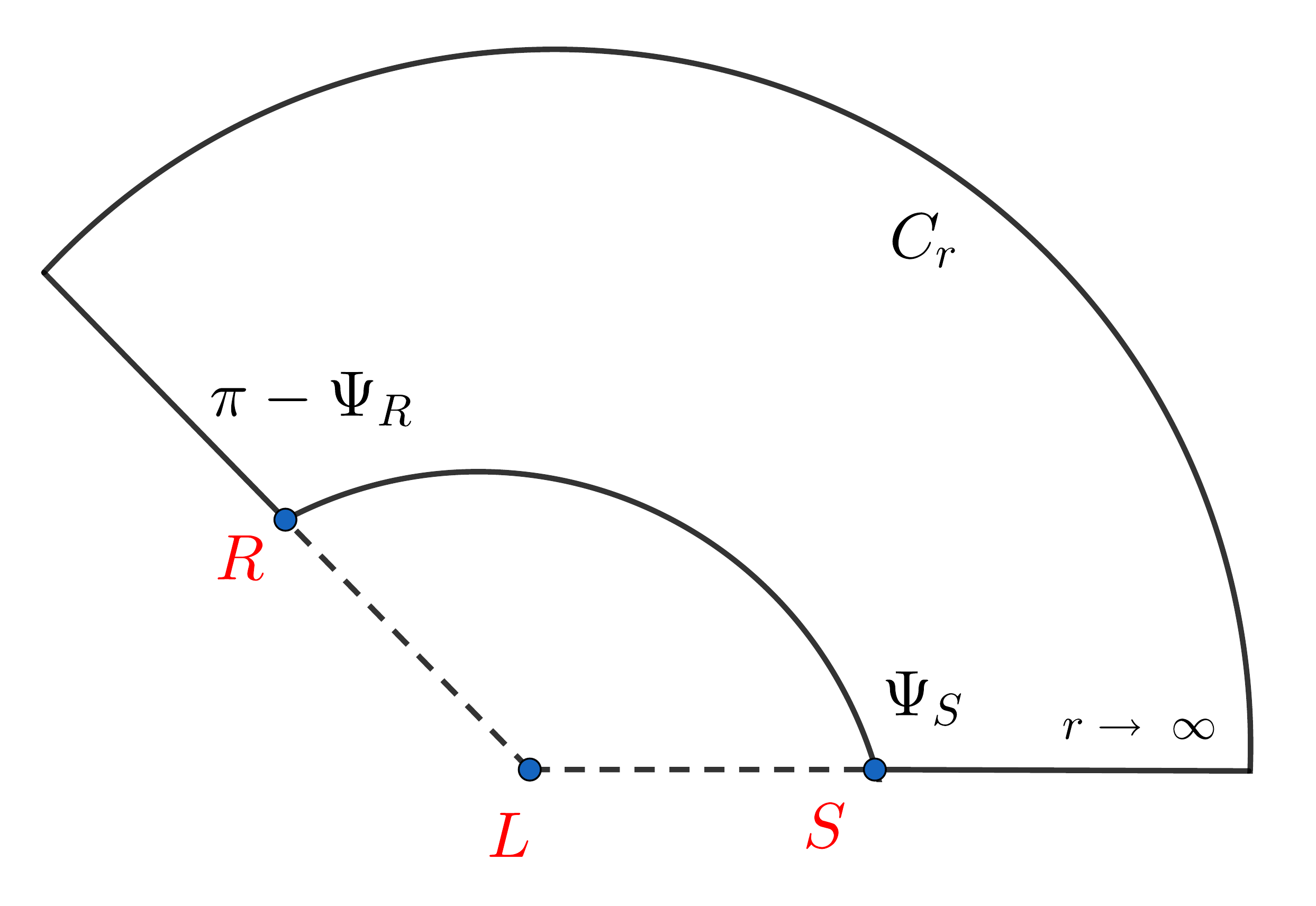}
  \caption{ The figure shows the  quadrilateral $_{R}^{\infty}\Box_{S}^{\infty}$ domain of integration. }
  \label{wedge}
  \end{figure}

 Next, let $\pi-\Psi_R$ and $\Psi_S$ be the corresponding inner angles measured at the vertices $R$ and
$S$, and consider a quadrilateral domain  $_{R}^{\infty}\Box_{S}^{\infty}$ embedded in a curved space as can be seen from  figure~\ref{wedge}. The quadrilateral $_{R}^{\infty}\Box_{S}^{\infty}$ consists of the light ray, two outgoing radial lines from $R$ and from $S$ and a circular arc segment $C_r$ at a coordinate distance $r_C \,(r_C \to \infty)$  from the coordinate origin located at the lens  $L$ (see  \cite{ish1,ish2,ish3} for more details).
Moreover, let $\phi_R$ and $\phi_S$ be the longitudes of the $R$
and the $S$, then we can define the quantity $\phi_{RS} \equiv \phi_R-\phi_S$, which gives the coordinate separation angle
between $R$ and $S$. By construction it follows that one can find a general relation for the deflection angle given in terms of  $\Psi_R$, $\Psi_S$ and $\phi_{RS}$,  by the following compact form \cite{ish1,ish2,ish3}
\begin{eqnarray}
\hat{\alpha}\equiv \Psi_R-\Psi_S+\phi_{RS}.
\end{eqnarray}

That being said, basically one can find the deflection angle $\hat{\alpha}$ by just computing  $\Psi_R$, $\Psi_S$ and $\phi_{RS}$ and applying the last relation. Note that this method will be used later on in the case of non-vanishing $\Lambda$.  There is, however, another way to find $\hat{\alpha}$ given the Gaussian curvature $\mathcal{K}$ and geodesic curvature $\kappa_g$. To do this, one simply has to integrate the Gaussian curvature  $\mathcal{K}$ over the quadrilateral $_{R}^{\infty}\Box_{S}^{\infty}$ domain. We recall that for an asymptotically flat
spacetime  $\kappa_g \to 1/r_C$ and $dl \to r_c d \phi $  as $r_c \to \infty $, implying the relation $\int_{C_r}\kappa_g dl \to \phi_{RS}$. Taking into account this information the GBT can be rewritten as follows \cite{ish1,ish2,ish3}
\begin{eqnarray}
\hat{\alpha}= -\iint_{_{R}^{\infty}\Box_{S}^{\infty}} \mathcal{K} dS+\int_S^R \kappa_g dl.
\end{eqnarray}

In what follows we are going to use this particular form of the GBT to calculate the finite distance corrections on the deflection angle of light. Considering the deflection of light in the equatorial plane and applying the definition of the
Gaussian curvature we find the following result for $\mathcal{K}$ in leading order
\begin{eqnarray}\notag
\mathcal{K}&=&\frac{R_{r \phi r \phi}}{\det \gamma }\\\notag
&=& \frac{1}{\sqrt{\det \gamma}}\left[\frac{\partial}{\partial \phi}\left(\frac{\sqrt{\det \gamma}}{\gamma_{rr}} \Gamma^{\phi}_{rr}  \right) -\frac{\partial}{\partial r}\left(\frac{\sqrt{\det \gamma}}{\gamma_{rr}} \Gamma^{\phi}_{r\phi}  \right)  \right]\\
&=& -\frac{2 M}{r^3}-\frac{\alpha \left(3-2 \log(\frac{r}{|\alpha |}\right)}{2r^3}+ {\mathcal{O}\left(\frac{M \alpha}{r^4},\frac{a^2 M}{r^5}\right)},
\end{eqnarray}
 in both $M$ and $\alpha$.

On the other hand, the geodesic curvature of the light ray for the stationary spacetime can be obtained by  \cite{ish3,ish4}
\begin{eqnarray}
\kappa_g=-\sqrt{\frac{1}{\det\gamma\, \gamma^{\theta \theta}}}\,\beta_{\phi,r}.
\end{eqnarray}
Using this last equation  we obtain the following result
\begin{eqnarray}
\kappa_g&=&-\frac{1}{\sqrt{\frac{r^8 \sin^2 \theta \left(r^2+\cos^2 \theta a^2\right)^2}{\left(\Delta_r-a^2 \sin^2 \theta\right)^4}\left(\frac{\Delta_r-a^2 \sin^2 \theta}{r^4}  \right)}}\,\beta_{\phi,r} \\
&=& -\frac{2 a M}{r^3}+\mathcal{O}\left(\frac{a \,\alpha}{r^4} \left[(2M+r) \log(\frac{r}{|\alpha|})-(M+r)\right]\right),
\end{eqnarray}
for the leading order term.

We can proceed to find the deflection angle by evaluating first the integration of $\mathcal{K}$ over the quadrilateral $_{R}^{\infty}\Box_{S}^{\infty}$ in terms of the following integral
\begin{eqnarray}\notag
-\iint_{_{R}^{\infty}\Box_{S}^{\infty}} \mathcal{K} dS &=&\int_{\phi_{S}}^{\phi_{R}} \int_{\infty}^{r(\phi)} \left(  -\frac{2 M}{r^3}-\frac{\alpha \left(3-2 \log(\frac{r}{|\alpha |}\right)}{2r^3} \right) \sqrt{\det \gamma}dr d \phi+ {\mathcal{O}\left(\frac{M \alpha}{b^2},\frac{a^2 M}{b^3}\right)}\\\notag
&=& \int_{\phi_{S}}^{\phi_{R}} d\phi\int_{0}^{\frac{\sin \phi}{b}} \left[2 M+\frac{3 \alpha}{2} - \alpha \log\left(\frac{1}{|\alpha| u}\right) \right]du \\\notag
&=& \int_{\phi_{S}}^{\phi_{R}}\left( \frac{ 2M \sin \phi}{b}+\frac{\alpha \sin \phi }{2 b}\right) d\phi+\int_{\phi_{S}}^{\phi_{R}}\left( -\frac{\sin \phi \, \alpha}{b} \log\left(\frac{b}{|\alpha| \sin \phi}\right)\right) d\phi\\\notag
&=& \frac{2M}{b}\left( \sqrt{1-b^2 u_R^2 }+\sqrt{1-b^2 u_S^2 }\right)+\frac{\alpha}{2b}\left( \sqrt{1-b^2 u_R^2 }+\sqrt{1-b^2 u_S^2 }\right)\\
&-&\frac{\alpha}{b}\left( (\sqrt{1-b^2 u_R^2 }+\sqrt{1-b^2 u_S^2 })(1+\log(\frac{b}{|\alpha|})-2 \log 2 \right) + {\mathcal{O}\left(\frac{M \alpha}{b^2},\frac{a^2 M}{b^3}\right)}
\end{eqnarray}
 where we have considered only the leading terms in $M$ and $\alpha$, and the light ray equation $r(\phi)=b/\sin(\phi)$ as the zeroth approximation of the deflected light ray.  Note that $b$ is the impact parameter, defined as $b \equiv \xi=L/E$, with $E$ being the energy of the particle (photon), and $L$ being the angular momentum of the particle measured at infinity.  Furthermore, we have used the relations $\cos \phi_{S}=  \sqrt{1-u_{S}^2 b^2}+{\mathcal{O}(Mu_S,au_S,\alpha u_S)}$ and $\cos \phi_R=-\sqrt{1-u_R^2 b^2}+ {\mathcal{O}(Mu_R,au_R,\alpha u_R)}$.  Note that in the above integral we have introduced a new variable $u=r^{-1}$, which is related to the finite radial distance of the source (receiver) from the black hole as follows $u_{S,R}=r_{S,R}^{-1}.$
The integral of $\kappa_g$ can be evaluated easily, yielding
\begin{eqnarray}\notag
\int_S^R \kappa_g dl &=&\int_S^R \left(-\frac{2Ma}{r^3}+{\mathcal{O}(\frac{\alpha M a}{r^4},\frac{a M^2}{r^4}})\right) dl\\\notag
&=& - \frac{2 Ma}{b^2}  \int_{\phi_{S}}^{\phi_{R}} \left( \cos \vartheta d\vartheta \right)+{\mathcal{O}(\frac{\alpha M a}{b^3},\frac{a M^2}{b^3})}\\\notag
&=& - \frac{2 Ma}{b^2} \left( \sin \phi_R-\sin \phi_S \right)+{\mathcal{O}(\frac{\alpha M a}{b^3},\frac{a M^2}{b^3})}\\
&=& \frac{2 Ma}{b^2} \left( \sqrt{1-b^2 u_R^2 }+\sqrt{1-b^2 u_S^2 }\right)+{\mathcal{O}(\frac{\alpha M a}{b^3},\frac{a M^2}{b^3})}.
\end{eqnarray}

 By adapting a coordinate system located at the lens $L$ one can parameterize the light equation in terms of the variable $\vartheta$;  thus by construction, it follows the light ray orbit  $r=b/\cos \vartheta+ \mathcal{O}(M,a,\alpha) $, with  $l=b \tan \vartheta+\mathcal{O}(M,a,\alpha)$ and $\sin \phi_{R,S}=  {\mp} \sqrt{1-u_{R,S}^2 b^2}+ {\mathcal{O}(Mu_{{R},S},au_{{R},S},\alpha u_{{R},S})}$.  Finally putting together these results we obtain the deflection angle
\begin{eqnarray}\notag
\hat{\alpha}&=& -\iint_{_{R}^{\infty}\Box_{S}^{\infty}} K dS+\int_{S}^{R} \kappa_g dl\\\notag
&=& \frac{2M}{b}\left( \sqrt{1-b^2 u_R^2 }+\sqrt{1-b^2 u_S^2 }\right)+\frac{\alpha}{2 b} \left( \sqrt{1-b^2 u_R^2 }+\sqrt{1-b^2 u_S^2 }\right) \\
&-&\frac{\alpha}{b} \zeta +  \frac{2 Ma}{b^2} \left( \sqrt{1-b^2 u_R^2 }+\sqrt{1-b^2 u_S^2 }\right)+ {\mathcal{O}\left(\frac{M \alpha}{b^2},\frac{a^2 M}{b^3},\frac{a M^2}{b^3},\frac{\alpha M a}{b^3}\right)},\label{def1}
\end{eqnarray}
where
\begin{eqnarray}
\zeta&=&  \left(\sqrt{1-b^2 u_R^2 }+\sqrt{1-b^2 u_S^2 }\right)\left(1+\log\left[\frac{b}{|\alpha|}\right]\right)-2 \log 2 .
\end{eqnarray}

This result shows that the Kerr deflection angle is  strongly affected by the PFDM parameter under the effect of finite distance corrections. In particular we see that the deflection angle is proportional to the PFDM parameter which belongs in the interval $ \alpha\in (-7.18M,0)\cup(0,2M)$. In the above expression we have evaluated the integral of the geodesic curvature $\kappa_g$ from  $S$ to $R$. Of course, one can evaluate this integral from $R$ to $S$ with a corresponding negative sign before the integral. In the general expression for the deflection angle, however, one should add the $\pm$ in order to include both cases.

Finally if we assume that $S$ and $R$ are located at  null infinity, i.e., $u_S \to 0$ and $u_R \to 0$, yielding  $\zeta =2 (1+\log(b/|\alpha|))-2 \log 2 $ resulting in the deflection angle
\begin{eqnarray}
\hat{\alpha} \to \frac{4M}{b}-\frac{\alpha}{b}\left(1-2 \log 2+ 2\log(\frac{b}{|\alpha|})   \right)\pm \frac{4  M a }{b^2}.
\end{eqnarray}

As a special case we can find the Kerr deflection angle by setting $\alpha=0$. Note that the $\pm $ sign corresponds for the retrograde and prograde light ray, respectively.
\begin{figure}[h!]
 \includegraphics[width=0.49\textwidth]{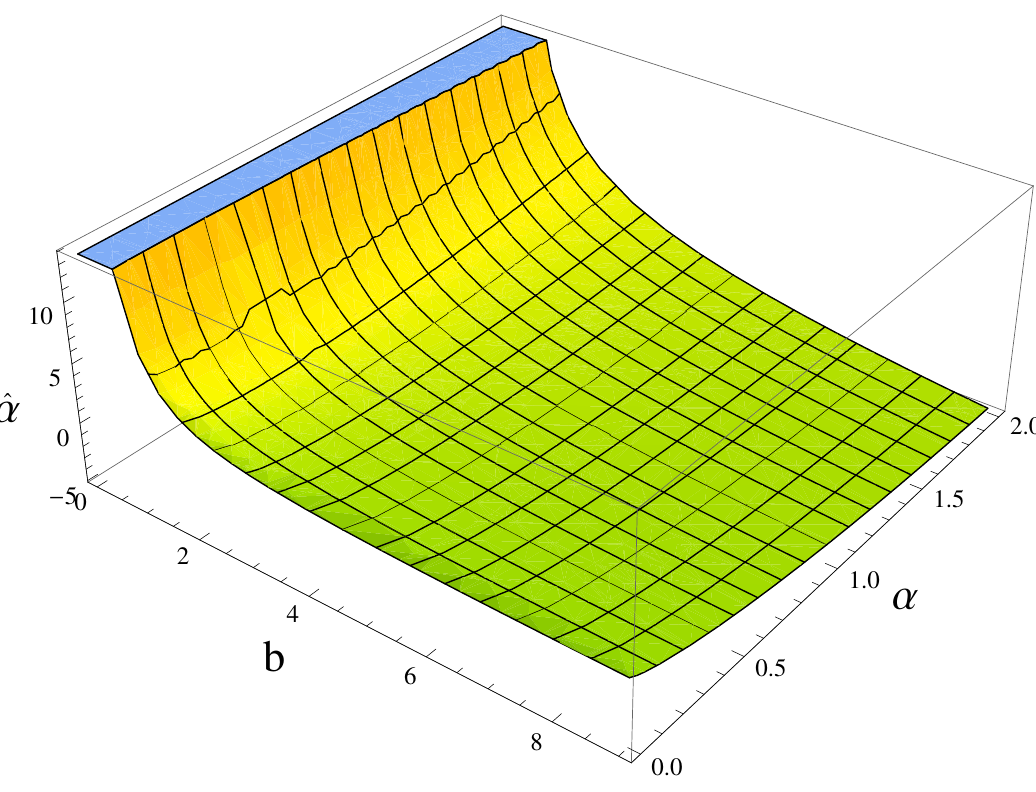}
 \includegraphics[width=0.49\textwidth]{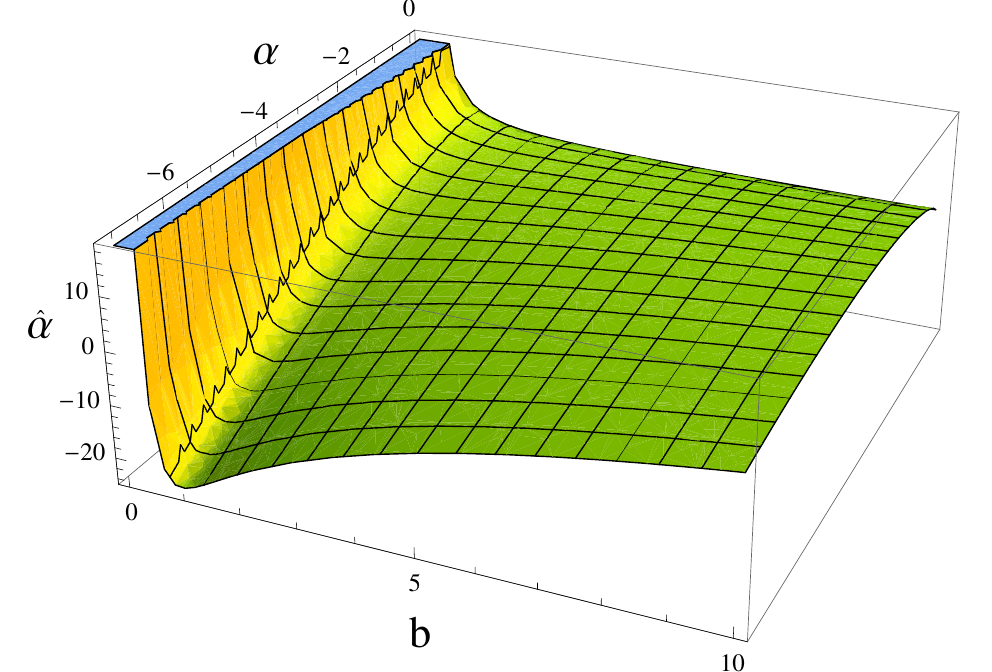}
  \caption{On the left side we plot the deflection angle $\hat{\alpha}$ as a function of the impact factor $b$ and the PFDM parameter $\alpha \in (0,2M)$. On the right side we plot $\hat{\alpha}$ as a function of the impact factor $b$ and the PFDM parameter $\alpha \in (-7.18  M,0)$. We have chosen $M=1$ and $a=0.8$ in both plots. }\label{fig3}
  \end{figure}

\subsection{Deflection angle with a cosmological constant}
Let us proceed by considering a more general scenario, namely by including the cosmological constant. In addition we are going to study the orbit equation on the equatorial plane.  The Lagrangian from the metric \eqref{metric} is found to be
%\begin{eqnarray}
%2\,\mathcal{L}=-\frac{\Delta_r (r)}{\Xi^2 r^2}(\dot{t}-a \dot{\phi})^2+\frac{1}{\Xi^2 r^2}(a \dot{t}-(r^2+a^2)\dot{\phi})^2+\frac{r^2}{\Delta_r(r)}\dot{r}^2
%\end{eqnarray}
\begin{equation}
2\,\mathcal{L}=-\mathcal{A}(r)\,\dot{t}^2-2 \mathcal{H}(r) \,\dot{t}\,\dot{\phi}+\mathcal{B}(r)\,\dot{r}^2+\mathcal{D}(r)\, \dot{\phi^2},
\end{equation}
where dot represents a derivation to the affine parameter $\lambda$. Furthermore we have introduced the following relations
\begin{eqnarray}
\mathcal{A}(r) &=& \left(\frac{\Delta_r-a^2}{\Xi^2 r^2}\right),\\
\mathcal{B}(r) &=& \frac{r^2}{\Delta_r },\\
\mathcal{H}(r) &=&  \frac{ar^2+a^3 -\Delta_r a}{\Xi^2 r^2},\\
\mathcal{D}(r) &=& \frac{r^4+2 a^2 r^2+a^4 -\Delta_r a^2 }{\Xi^2 r^2}.
\end{eqnarray}

Due to the spacetime symmetries there are two Killing vectors associated with the metric \eqref{metric}, implying two constants of motion defined  as follows
\begin{eqnarray}
-p_t=\frac{\partial \mathcal{L}}{\partial \dot{t}}=E=\mathcal{A}(r) \dot{t}+\mathcal{H}(r)\dot{\phi},
\end{eqnarray}
and
\begin{eqnarray}
p_{\phi}=\frac{\partial \mathcal{L}}{\partial \dot{\phi}}=L=-\mathcal{H} \dot{t}+\mathcal{D}(r)\dot{\phi}.
\end{eqnarray}

The light ray equation can be obtained by defining first the impact parameter as
\begin{eqnarray}
b=\frac{L}{E}=\frac{\mathcal{D}(r) \frac{d \phi}{dt}-\mathcal{H}(r)}{\mathcal{H}(r) \frac{d \phi}{dt}+\mathcal{A}(r)},
\end{eqnarray}
then by introducing a new variable as $u=1/r$, with the following important relation
\begin{equation}
\frac{\dot{r}}{\dot{\phi}}=\frac{dr}{d \phi}=-\frac{1}{u^2}\frac{d u}{d \phi}.
\end{equation}

Then the light ray equation is given by the differential equation
\begin{eqnarray}
\left(\frac{d u}{d \phi}\right)^2=\frac{u^4 \left(\mathcal{A}(u)\mathcal{D}(u)+\mathcal{H}^2(u)   \right)  \left( \mathcal{D}(u)-2 \mathcal{H}(u) b-\mathcal{A}(u) b^2 \right)}{\mathcal{B}(u) \left(\mathcal{H}(u)-\mathcal{A}(u) b  \right)^2}\equiv F(u).
\end{eqnarray}

From the GBT  we recall the following relation for the deflection angle
\begin{eqnarray}\notag
\hat{\alpha} & \equiv & \Psi_R-\Psi_S+\phi_{RS}\\
&=& \Psi_R-\Psi_S+\int_{u_R}^{u_0}\frac{du}{\sqrt{F(u)}}+\int_{u_S}^{u_0}\frac{du}{\sqrt{F(u)}},\label{eq59}
\end{eqnarray}
%where the inverse of the closest approach can be approximated with the impact factor i.e., $u_0=b$.
 where the closest approach $r_0=u_0^{-1}$ (which is simply the closest distance of the photon to the black hole) in the weak limit can be approximated with the impact parameter $b$ i.e. $b=u_0^{-1}$.  Using the unit tangent vector  $e_i$ along the light ray orbit in the manifold $\mathcal{M}$ which satisfies the relation $\gamma_{ij}e^i e^j=1$, it is possible obtain  the angles at the $R$ and $S$ in terms of the following quantity \cite{ish1}
\begin{eqnarray}\label{eq60}
\sin \Psi=\frac{\mathcal{H}(r)+\mathcal{A}(r) b}{\sqrt{\mathcal{A}(r) \mathcal{D}(r)+\mathcal{H}^2(r)}}.
\end{eqnarray}

From  Eq. \eqref{eq60} it is possible to determine the quantity $\Psi_R-\Psi_S$ which in our case is found to be
\begin{eqnarray}\notag
\Psi_R-\Psi_S &=&\Psi^{Kerr}_R-\Psi^{Kerr}_S-\frac{b \Lambda }{6}\left( \frac{1}{u_R \sqrt{1-b^2 u_R^2 }}+\frac{1}{u_S \sqrt{1-b^2 u_S^2 }}\right)\\\notag
&+&\frac{b\Lambda M}{6}\left(\frac{2 b^2 u_R^2-1}{(1-b^2 u_R^2)^{3/2} }+\frac{2 b^2 u_S^2-1}{(1-b^2 u_S^2)^{3/2}}\right)\\\notag
&+&\frac{a\Lambda}{3}\left(\frac{1}{u_R\sqrt{1-b^2 u_R^2 }}+\frac{ 1 }{u_S \sqrt{1-b^2 u_S^2 }}\right)\\\notag
&+& \frac{\alpha b}{2}\left(\frac{ u_R^2  \log(\frac{1}{u_R |\alpha|})}{ \sqrt{1-b^2 u_R^2 } }+\frac{u_S^2  \log(\frac{1}{u_S |\alpha|})}{ \sqrt{1-b^2 u_S^2 } }\right)\\
%&+& \alpha a \left(\frac{ u_R^2  \log(\frac{1}{u_R |\alpha|})}{ \sqrt{1-b^2 u_R^2 } }+\frac{u_S^2  \log(\frac{1}{u_S |\alpha|})}{ \sqrt{1-b^2 u_S^2 } }\right)\\
%&+& \frac{\alpha M b}{2} \left[\frac{u_R^3 \log(\frac{1}{u_R |\alpha|})(2 b^2u_R^2-1)}{(1-b^2 u_R^2)^{3/2}} +\frac{u_S^3 \log(\frac{1}{u_S |\alpha|})(2 b^2u_S^2-1)}{(1-b^2 u_S^2)^{3/2}}   \right]\\
&-& \frac{\alpha \Lambda b}{12}\left( \frac{(2 b^2u_R^2-1)}{(1-b^2 u_R^2)^{3/2}}+\frac{(2 b^2 u_S^2-1)}{(1-b^2 u_S^2)^{3/2}} \right)+{\mathcal{O}(a\Lambda M,\alpha M/b,\alpha a/b^2, M^2 \Lambda,\alpha^2 \Lambda)},
\end{eqnarray}
where
\begin{eqnarray}\notag
\Psi^{Kerr}_R-\Psi^{Kerr}_S&=&(\arcsin(b u_R)+\arcsin(b u_S)-\pi)\\\notag
&-& b M\left(\frac{u_R^2}{\sqrt{1-b^2 u_R^2}}+\frac{u_S^2}{\sqrt{1-b^2 u_S^2}}\right) \\
& +&2 a M \left(\frac{u_R^2}{\sqrt{1-b^2 u_R^2}}+\frac{u_S^2}{\sqrt{1-b^2 u_S^2}}  \right).
\end{eqnarray}

Our next goal is to compute the quantity $\phi_{RS}$ in leading order terms by evaluating the integral of the angular coordinate $\phi$ in terms of the following equation
\begin{eqnarray}\notag
\phi_{RS} &=& \int_S^R d \phi \\\notag
&=& \phi_{RS}^{Kerr}-\frac{\alpha}{2 b}\left(\sqrt{1-b^2 u_R^2}+\sqrt{1-b^2 u_S^2}\right)-\frac{\alpha \Xi }{b}\\\notag
&-&\frac{\alpha}{b}\Big[\frac{(2-b^2 u_R^2)\log(\frac{1}{u_R |\alpha|})}{2\sqrt{1-b^2 u_R^2 }}+\frac{(2-b^2 u_S^2)\log(\frac{1}{u_S |\alpha|})}{2\sqrt{1-b^2 u_S^2 }} \Big]\\\notag
&+&\frac{\Lambda b^3}{6}\left(\frac{u_R}{\sqrt{1-b^2 u_R^2}} +\frac{u_S}{\sqrt{1-b^2 u_S^2}}  \right)\\\notag
&+& \frac{M\Lambda b}{6}\left[\frac{2-3 b^2 u_R^2}{(1-b^2 u_R^2)^{3/2}} +\frac{2-3 b^2 u_S^2}{(1-b^2 u_S^2)^{3/2}}    \right]\\\notag
&+& \frac{a \Lambda }{3}\left(\frac{1-2 b^2 u_R^2}{u_R\sqrt{1-b^2 u_R^2}} +\frac{1-2b^2 u_S^2}{u_S\sqrt{1-b^2 u_S^2}}  \right)\\
&+& \frac{\alpha \Lambda b}{12}\left( \frac{1}{\sqrt{1-b^2 u_R^2}} +\frac{1}{\sqrt{1-b^2 u_S^2}} \right)+{\mathcal{O}(a\Lambda M,\alpha M/b,\alpha a/b^2, M^2 \Lambda,\alpha^2 \Lambda)},
\end{eqnarray}
where we have introduced
\begin{eqnarray}\notag
\phi_{RS}^{Kerr}&=& \pi-(\arcsin(b u_R)-\arcsin(b u_S))\\\notag
&+& \frac{2 M}{b}\left[ \frac{1}{\sqrt{1-b^2 u_R^2}}\left(1-\frac{1}{2}b^2 u_R^2  \right)+\frac{1}{\sqrt{1-b^2 u_S^2}}\left(1-\frac{1}{2}b^2 u_S^2  \right)   \right]\\
&-& \frac{2a M}{b^2}\left(\frac{1}{\sqrt{1-b^2 u_R^2}}+\frac{1}{\sqrt{1-b^2 u_S^2}}\right),
\end{eqnarray}
and
\begin{equation}
\Xi=2 \log(b)-\log[(1+\sqrt{1-b^2 u_R^2 })(1+\sqrt{1-b^2 u_S^2 })].
\end{equation}

Going back to Eq. \eqref{eq59} and after some algebraic manipulations we obtain the following result for the deflection angle
\begin{eqnarray}\notag
\hat{\alpha}&=& \frac{2M}{b}\left( \sqrt{1-b^2 u_R^2 }+\sqrt{1-b^2 u_S^2 }\right)-\frac{\alpha}{2 b} \left( \sqrt{1-b^2 u_R^2 }+\sqrt{1-b^2 u_S^2 }\right) \\\notag
&-&\frac{\alpha}{b} \Big[\Xi+(\sqrt{1-b^2 u_R^2 }+\sqrt{1-b^2 u_S^2 })\log(\frac{1}{|\alpha|})\Big] \\\notag
&-&\frac{\Lambda b}{6 }\left[ \frac{\sqrt{1-b^2 u_R^2 }}{u_R  }+\frac{\sqrt{1-b^2 u_S^2 }}{u_S  }  \right]\\\notag
&+ & \frac{M \Lambda b}{6 }\left[ \frac{1}{ \sqrt{1-b^2 u_R^2 } }+\frac{1}{\sqrt{1-b^2 u_S^2 } }  \right]\\\notag
&+& \frac{\alpha \Lambda b}{12 }\left[ \frac{2-3b^2 u_R^2}{ (1-b^2 u_R^2)^{3/2} }+\frac{2-3b^2 u_S^2}{(1-b^2 u_S^2)^{3/2} }  \right]\pm \frac{2 a\Lambda}{3}\left[\frac{\sqrt{1-b^2 u_R^2 }}{u_R}  +\frac{\sqrt{1-b^2 u_S^2 }}{u_S} \right] \\
&\pm & \frac{2 Ma}{b^2} \left( \sqrt{1-b^2 u_R^2 }+\sqrt{1-b^2 u_S^2 }\right)+{\mathcal{O}(a\Lambda M,\alpha M/b,\alpha a/b^2, M^2 \Lambda,\alpha^2 \Lambda)}.
\end{eqnarray}

In the case  of a vanishing cosmological constant we recover Eq. \eqref{def1}.
Although there are terms that diverge in the case of a nonzero cosmological constant (in the limit $bu_R \to 0$ and $bu_S \to 0$), from a physical point of view we know that an observed star or galaxy is located at a finite distance from us.  In other words, the limit $bu_R \to 0$ and $bu_S \to 0$, is not allowed in this case but one can include only a certain finite distance which leads to further simplified relation
\begin{eqnarray}\notag
\hat{\alpha}&\sim & \frac{4M}{b}-\frac{\alpha}{b}\left(1-2 \log 2+ 2\log(\frac{b}{|\alpha|})   \right) -\frac{\Lambda b}{6 }\left( \frac{1}{u_R  }+\frac{1}{u_S } \right)+  \frac{M \Lambda b}{3 }+ \frac{\alpha \Lambda b}{3 } \\
&\pm & \frac{4 Ma}{b^2} \pm \frac{2 a \Lambda}{3}\left( \frac{1}{u_R  }+\frac{1}{u_S } \right).
\end{eqnarray}

Finally we see that, besides the effect of PFDM parameter $\alpha$, there are additional corrections including a finite contribution term $\sim \alpha \Lambda b/3$, and a divergent term $a \Lambda/(u_R^{-1}+u_S^{-1})$. Here the limit $bu_R \to 0$ and $bu_S \to 0$ is not allowed -- in other words, we can consider only finite distance corrections. In this sense, the last equation generalizes a previous result reported in Ref. \cite{ish1}.

\section{Conclusion}

The existence of dark matter around black holes located at
the centers of most of large galaxies plays an important role in many astrophysical phenomena.
Motivated by this fact, in this paper we have studied the effects of  perfect fluid dark matter and a cosmological constant on the shadow of a rotating black hole.  Our work provides a possible tool for observation of
dark matter via  shadows, perhaps using the high resolution imaging of the Event Horizon Telescope.

We have shown that the different shadow shapes are found by
varying the PFDM parameter, mass, spin parameter, and the cosmological constant. Through graphs we have demonstrated that size of shadow of our black hole decreases for $\alpha <1$ with $M=1$ but after that we see an increase in its size. In addition we have done a detailed analyses on the effect of those parameters on the deflection angle of light using a recent geometric method by means of the the Gauss-Bonnet theorem applied to the optical geometry. Due to the presence of the cosmological constant we have included   finite distance correction on the deflection angle.

More specifically, in the case of  vanishing $\Lambda$, we found $\mathcal{K}$ and $\kappa_g$, then the deflection angle is simply found by integrating over the quadrilateral $_{R}^{\infty}\Box_{S}^{\infty}$ domain. Our results show that the PFDM parameter strongly affects the deflection of light under finite distance corrections.
 As a special case, in the limit  $bu_R \to 0$ and $bu_S \to 0$ the standard Kerr deflection angle is modified. In the case of a non-vanishing $\Lambda$ we followed an alternative method by computing the quantity $\Psi_R-\Psi_S+\phi_{RS}$ which gives the deflection angle. Besides $\alpha$, here we found a finite contribution term $\sim \alpha \Lambda b/3$, and a divergent term $a \Lambda/(u_R^{-1}+u_S^{-1})$. We pointed out that in this case the limit  $bu_R \to 0$ and $bu_S \to 0$ is not allowed. In this context, the presence of the divergent terms is not problematic due to the finite distance corrections. After all, by observations we can only observe a given star or a galaxy in finite distance from us.

 \section*{Acknowledgements}
This work was supported in part by the Natural Sciences and Engineering Research Council of Canada and National Natural Science Foundation of China (NNSFC) under contract No.11805166.

\end{document}